\documentclass[twoside]{dis09}
\usepackage[latin1]{inputenc}
\usepackage[dvips]{graphicx,epsfig,color}
\usepackage{wrapfig,rotating}
\usepackage{amssymb,amsmath,array}

\newcommand{\beq}{\begin{equation}}
\newcommand{\eeq}{\end{equation}}
\newcommand{\bac}{\beq\begin{array}}
\newcommand{\eac}{\end{array}\eeq}
\newcommand{\ba}{\begin{array}}
\newcommand{\ea}{\end{array}}
\newcommand{\bea}{\begin{eqnarray}}
\newcommand{\eea}{\end{eqnarray}}

\begin{document}
\title{DIS 2009 Concluding Talk: Outlook and Perspective}

\author{Guido Altarelli 
%
\thanks{Dedicated to the memory of Wu-Ki Tung, a leading scientist in the field of deep inelastic scattering.}
%
\vspace{.3cm}\\
%
Dipartimento di Fisica "E. Amaldi" and Sezione  INFN, Universita' di Roma Tre,
Roma, Italy\\%
and Department of Physics, Theory Unit, CERN, Geneva, Switzerland\\
}

\maketitle

\begin{abstract}
I will present here my perception on the status of Deep Inelastic Scattering physics, as I have further developed it during this Workshop, together with a number of comments on the results that have impressed me most during this week. I will emphasize a number of open problems and of critical areas. Finally I conclude with some projections and auspices for the future of the field.
\end{abstract}

\begin{flushright}
{RM3-TH/09-14}~~~~~~
{CERN-PH-TH/2009-114}\\
\end{flushright}

\section{Preamble}

Let me make clear from the start that I did not try to make a summary of the Workshop. The Convenors of the different working groups have presented their resumes just before my talk. Thus I felt free to leave aside many valuable contributions only because they do not fit in my chosen framework. For example, all reviews presented at the Workshop on the results obtained at hadron colliders or of studies in preparation for the LHC are not covered here. Rather I tried to write a personal view on the achievements and the open problems of Deep Inelastic Scattering as they appear now at a moment when the future of the field is not clear and the continuation of dedicated experiments is not yet firmly established.

\section{Forty years of tremendous progress}

Deep Inelastic Scattering (DIS) processes
have played and still play a very important role for our understanding of QCD and of nucleon structure. In the past DIS processes were crucial for establishing QCD as
the theory of strong interactions and for imposing quarks and gluons as the QCD partons. In the '60's the demise of hadrons from the  status of fundamental particles to that of bound states composed of constituent quarks was
the breakthrough that made possible the construction of a renormalisable field theory for strong interactions. The evidence for constituent quarks had already clearly emerged  from the systematics of hadron spectroscopy. But confinement that forbids 
the observation of free quarks was an obvious obstacle towards the acceptance of quarks as real constituents and not just as
fictitious entities describing some mathematical pattern (a doubt expressed even by Gell-Mann at the time). The early
measurements on DIS at SLAC  forty years ago \cite{slac} dissipated all doubts: the observation of approximate Bjorken scaling and the success of the "naive" (not so
much after all) parton model of Feynman imposed quarks as the basic fields for describing the nucleon structure. At present DIS remains very important for quantitative studies and tests of QCD: those based on scaling violations are among the most solid and comprehensive. The measurement of quark and
gluon densities in the nucleon is performed in DIS processes, as functions of x at some reference value of $Q^2$, which is an essential starting point, through the factorisation theorem, for
the calculation of all relevant hadronic hard processes. At the same time one measures
$\alpha_s(Q^2)$ and the DIS values of the running coupling can be compared with those obtained from other processes. Over the years new theoretical challenges have been derived from the study of DIS processes, like those related to the resummation of large logs, to the spin crisis and so on. 

\begin{figure}[!h]
\includegraphics[width=0.8\columnwidth]{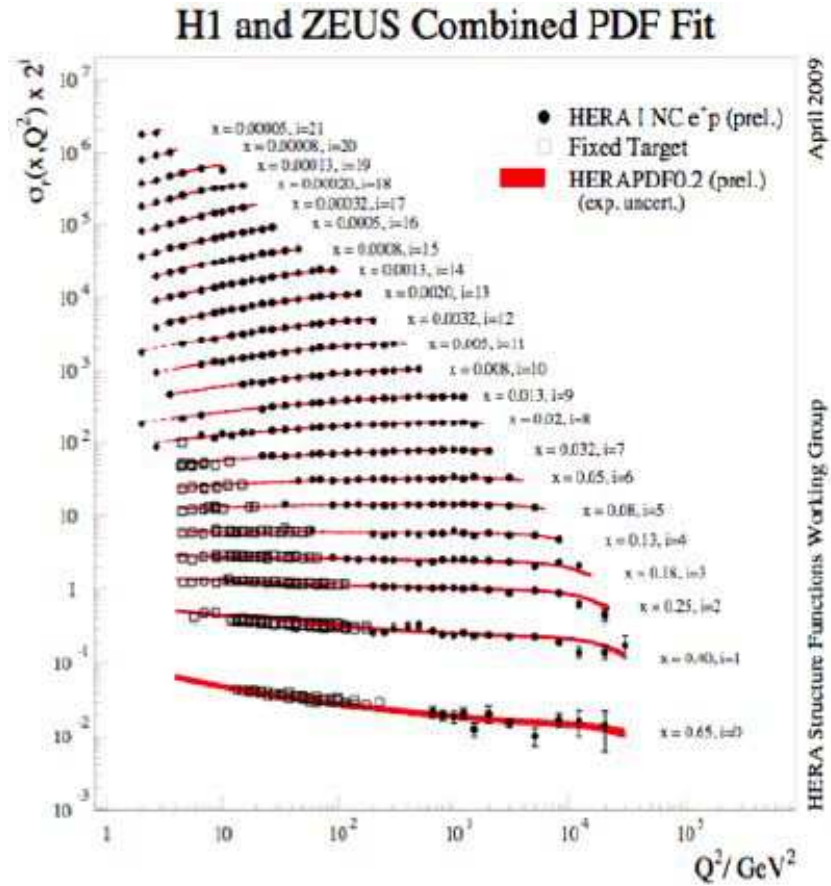} 
\caption[]{A recent NLO fit of scaling violations from the combined H1 and ZEUS data, for different $x$ ranges, as functions of $Q^2$.}
\label{fig1}
\end{figure}

\begin{figure}[!t]
\includegraphics[width=0.8\columnwidth]{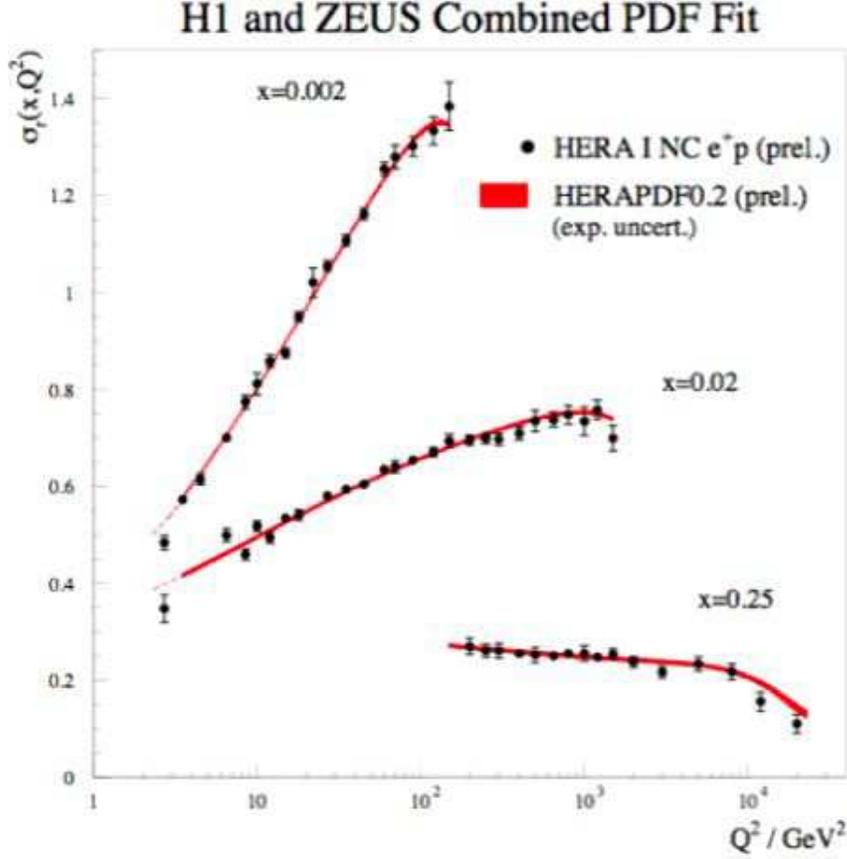} 
\caption[]{An example of the present status of scaling violations for DIS structure functions with a comparison to a NLO QCD fit.}
\label{Fig:SV}
\end{figure}

Formidable results on the physics of DIS have been obtained over the years both in experiment and in theory. Some of them are still being obtained and have been presented at this Workshop. As examples I will mention just a few items that I consider particularly impressive, but my list is by far not complete. As I mentioned, the first SLAC experiments forty years ago \cite{slac} established approximate scaling for structure functions and observed the dominance of the transverse cross section. In Figs. (\ref{fig1}, \ref{Fig:SV}) we see examples of how well the scaling violations are measured at present and the accurate description of the data by a Next-to-the-Leading Order (NLO) QCD fit (as illustrated in the talks by A. Cooper Sarkar, D. Gabbert, J. Kretzschmar, A. Petrukhin, V. Radescu and E. Tassi).  The present fits to scaling violations provide an impressive confirmation of the precisely quantitative QCD
predictions, a measurement of $q_i(x,Q_0^2)$ and $g(x,Q_0^2)$ at some reference value $Q_0^2$ and an accurate measurement of $\alpha_s(m_Z^2)$ (see later).

\begin{figure}[!t]
\includegraphics[width=0.8\columnwidth]{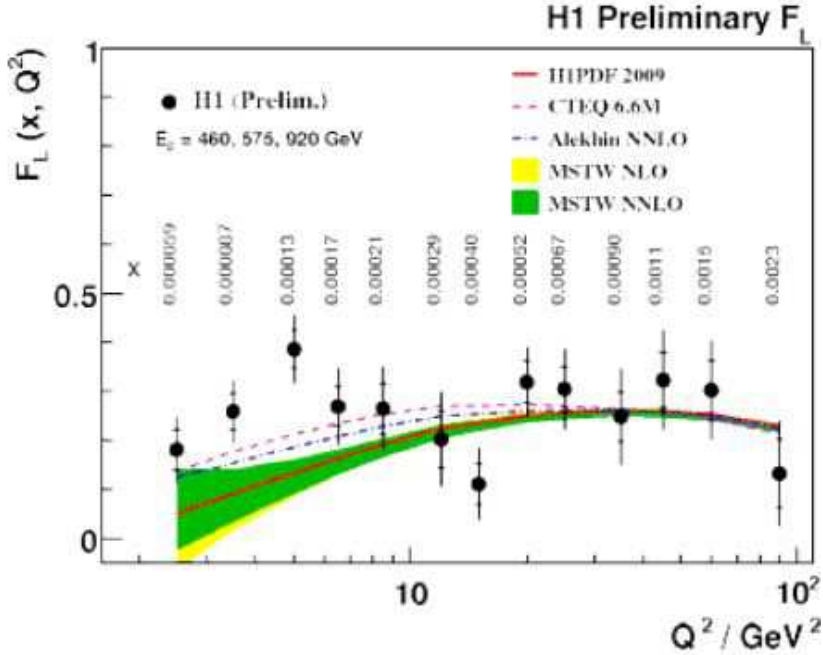} 
\caption[]{The longitudinal structure function $F_L$ measured by H1 at HERA.}
\label{longi}
\end{figure}

After SLAC established the dominance of the transverse cross section it took ~40 years to get meaningful data on the longitudinal structure function! These data represent an
experimental highlight of recent years (presented by S. Glazov, J. Grebenyuk, D. Salek, S. Schmitt). They have been obtained by H1 and ZEUS at HERA and it is a real pity that the data taking has been stopped. Theoretical aspects have been discussed in the talks by A. Stasto and D. Schildknecht. The data are shown in Fig.(\ref{longi}). In QCD at large $Q^2$ and at LO the simple, 30 years old, formula is valid (for $N_f=4$) \cite{am}:
\beq
F_L(x,Q^2)=\frac{\alpha_s(Q^2)}{2\pi}x^2\int_x^1\frac{dy}{y^3}\left[ \frac{8}{3}F_2(y,Q^2)+\frac{40}{9}yg(y,Q^2)(1- \frac{x}{y})\right]\\
\label{FL}
\eeq
I had not expected that it would take such a long time to have a meaningful test of this simple prediction! And in fact better data would be highly desirable. But how and when they will be obtained is not clear.

\begin{figure}[!h]
\includegraphics[width=0.8\columnwidth]{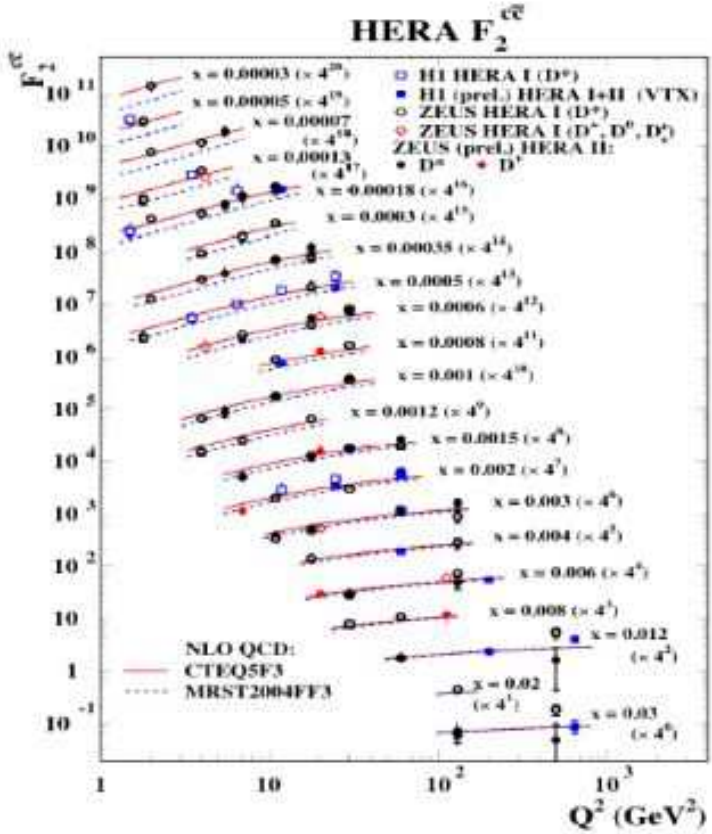} 
\caption[]{The charm structure function $F_2^{c \bar c}$ measured at HERA.}
\label{Fig:CSF}
\end{figure}

\begin{figure}[!t]
\includegraphics[width=0.8\columnwidth]{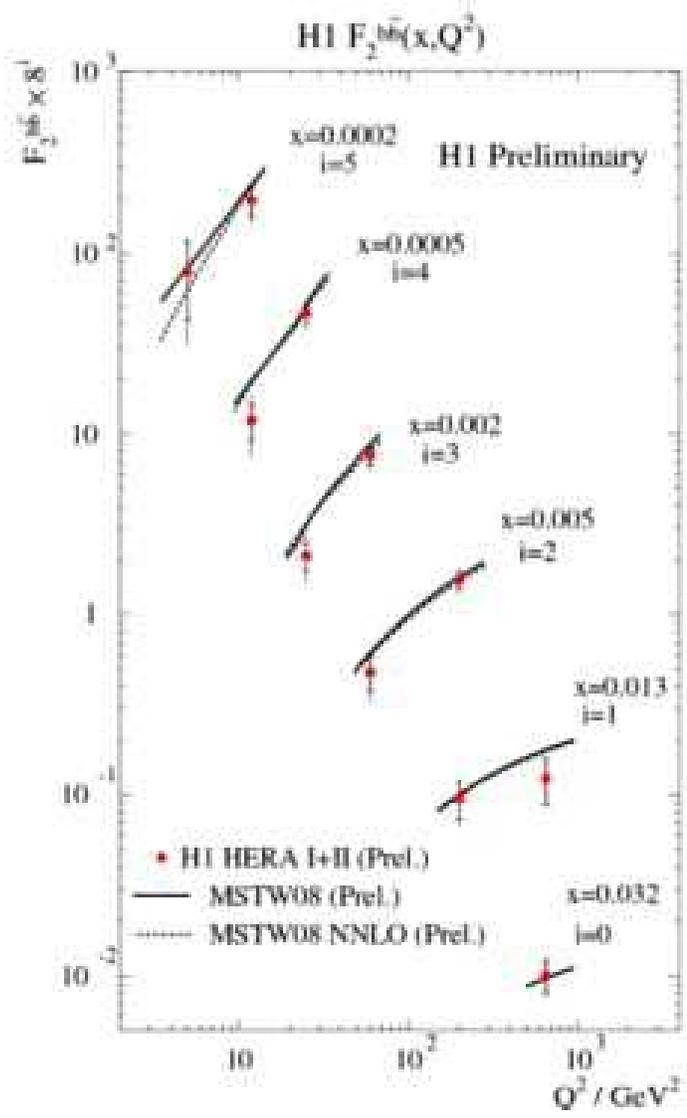} 
\caption[]{The bottom structure function $F_2^{b \bar b}$ measured at HERA.}
\label{Fig:BSF}
\end{figure}

\begin{figure}[!b]
\includegraphics[width=0.8\columnwidth]{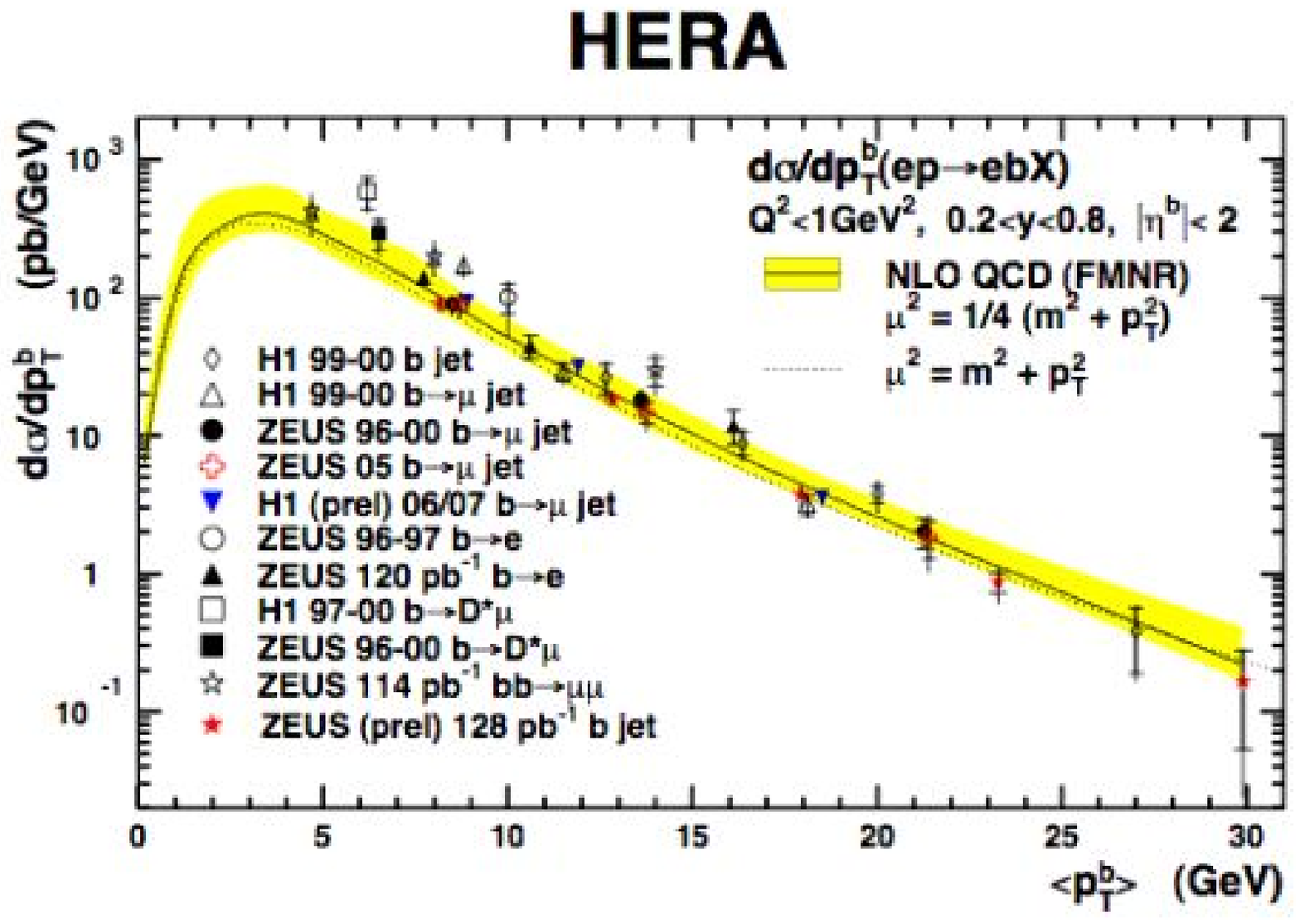} 
\caption[]{The b photoproduction crosssection compared to NLO QCD prediction.}
\label{bphoto}
\end{figure}

In recent years new domains of DIS physics have been opened and/or brought to maturity by experiment. For example, I was impressed by the remarkable progress done at HERA on heavy flavoured structure functions (presented by A. Jung, K. Lipka, P. Roloff, P. Thompson  and the convenor report by L. Gladilin and for related theory progress, S. Alekhin, S. Klein, T. Uematsu and the summary by A. Ali). Two examples, for charm and bottom, are shown in Figs. (\ref{Fig:CSF}, \ref{Fig:BSF}). We see that one has accomplished a quantitative description of the charm and bottom structure functions in $x$ and $Q^2$ with a clear pattern of scaling violations. I was also impressed by the good agreement now obtained for b photoproduction with NLO QCD predictions, as shown in Fig. \ref{bphoto} (talks by B. List, S. Miglioranzi, E. Tassi, T. Toll).

\begin{figure}[!t]
\includegraphics[width=0.8\columnwidth]{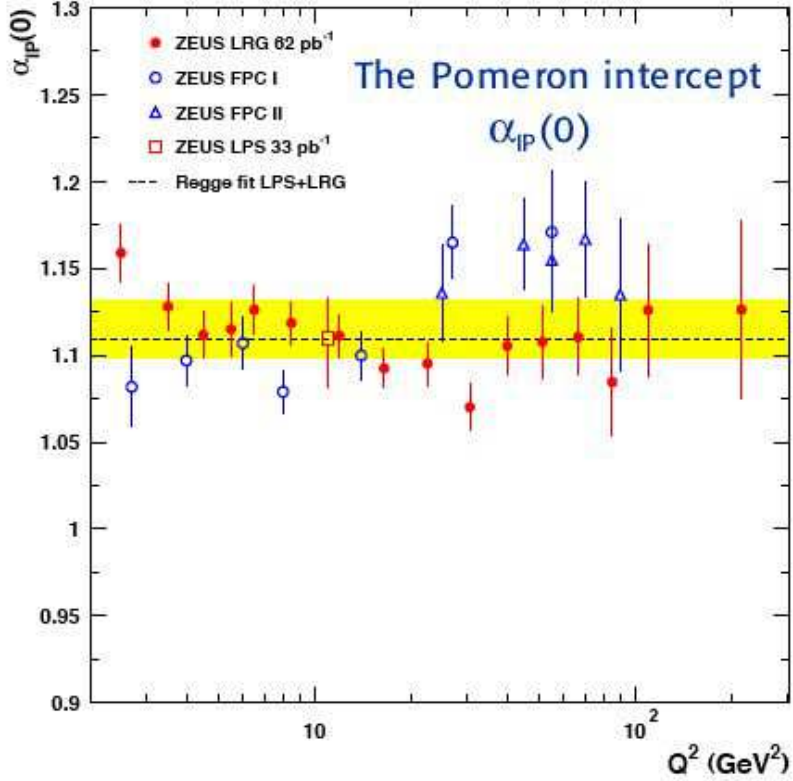} 
\caption[]{The "Pomeron" intercept measured at HERA.}
\label{pom}
\end{figure}

A set of really imposing experimental information is provided by the collection of data on diffractive structure functions (presented by M. Ruspa, W. Slominski, for theory see the talks by A. Luszczak, L. Schoeffel; see also the summaries by M. Diehl and J. P. Laycock ) that make quantitative our knowledge of what we could call the Pomeron structure functions, where by Pomeron (not a clarified concept in QCD) we denote whatever dominates the colour and gauge singlet t-channel at high energy. I found remarkable that so many aspects of the problem have consistently been explored by experiment and the various parts have been put together in a coherent picture: e.g. the measurement of the diffractive peak and its slope, the Regge fit, as function of $x_P$, the measurement of the Pomeron intercept resulting from the fit, $\alpha_P \sim 1.108 \pm 0.008 +0.022-0.007$ (Fig.\ref{pom}) which, interestingly, is found larger than 1 (perhaps 1 modulo logs). A lot of data are ready for more theoretical studies!

An interesting collection of data has been accumulated on exclusive production of vector mesons both from virtual photons and in photoproduction (discussed by A. Levy, P. Marage and  S. Yashenko). In particular for J/? production at HERA a straightforward NLO QCD colour singlet model fails in the predictions of both rates and polarization (see the talks by P. Artoisenet, A. Bertolin and P. Faccioli). Should we worry? Probably not too much, but it is an interesting problem to study. 

Progress in experiment has been matched by outstanding theory accomplishments. We already said that the theory
of scaling violations for totally inclusive DIS structure functions, based on operator expansion or diagrammatic techniques and on renormalisation group
methods, is crystal clear and the predicted $Q^2$ dependence can separately be tested for each structure function at each value of x. For many years the splitting functions were  completely known at NLO accuracy only: 
\beq
\alpha_s P~\sim~\alpha_s P_1~+~\alpha_s^2 P_2~+.... \\
\eeq
Beyond leading order a precise definition of parton densities should be specified.  Once the definition of parton densities is fixed, the coefficients that relate the different structure functions to the parton densities at each fixed order can be computed. As a consequence, the higher order splitting functions also depend, to some extent, from the definition of parton densities, and a consistent set of coefficients and splitting functions must be used at each order. Finally in recent years the NNLO results
$P_3$ have been first derived in analytic form for the first few moments and, then the full NNLO analytic calculation, a really monumental work, was completed in 2004 by Moch, Vermaseren and Vogt \cite{ver}. The coefficients in the same scheme, to the appropriate perturbative level, have also been computed \cite{coeff}. I am very admired of the accomplishments of the most advanced calculational techniques of present days (as, for example, reviewed in the talk by N. Glover) and of the outstanding skill of the people involved. At this Workshop recent results have been presented, for example, by A. Vogt on the $N^3LO$ coefficients for the $F_3$ structure function and by J. Bluemlein and J. A. M. Vermaseren on some computational advances. 

\begin{figure}[!h]
\includegraphics[width=0.8\columnwidth]{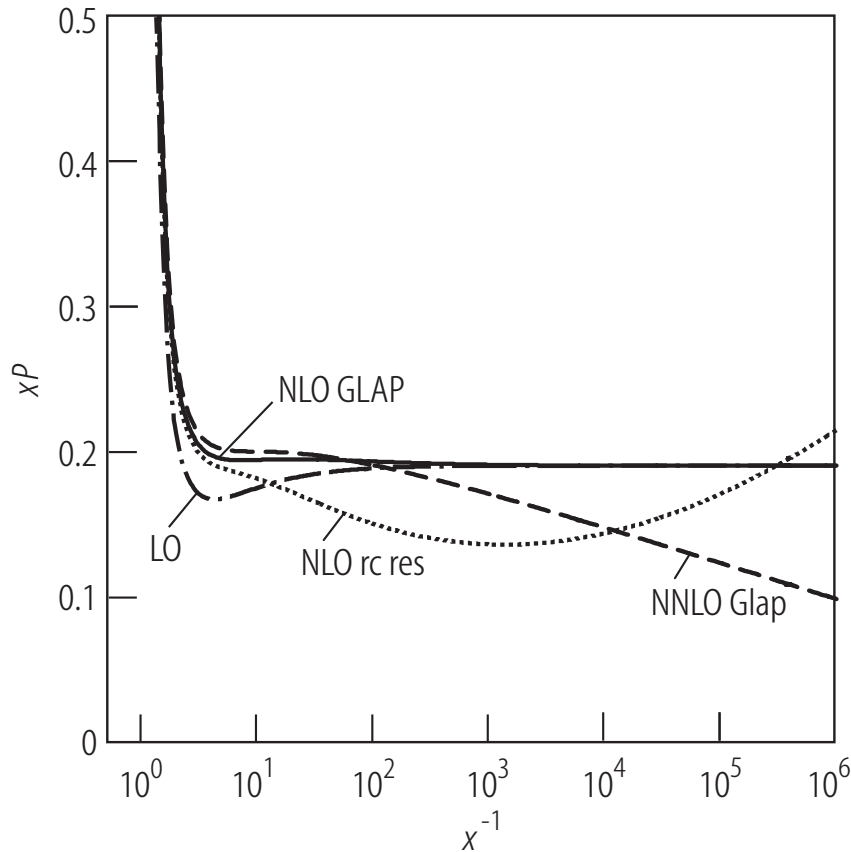} 
\caption[]{The dominant singlet splitting function $xP(x,\alpha_s(Q^2))$ for $\alpha_s(Q^2) \sim 0.2$. The resummed result is compared with the LO, NLO and NNLO perturbative results \cite{heralhc}.}
\label{figx}
\end{figure}

\begin{figure}[!h]
\includegraphics[width=0.8\columnwidth]{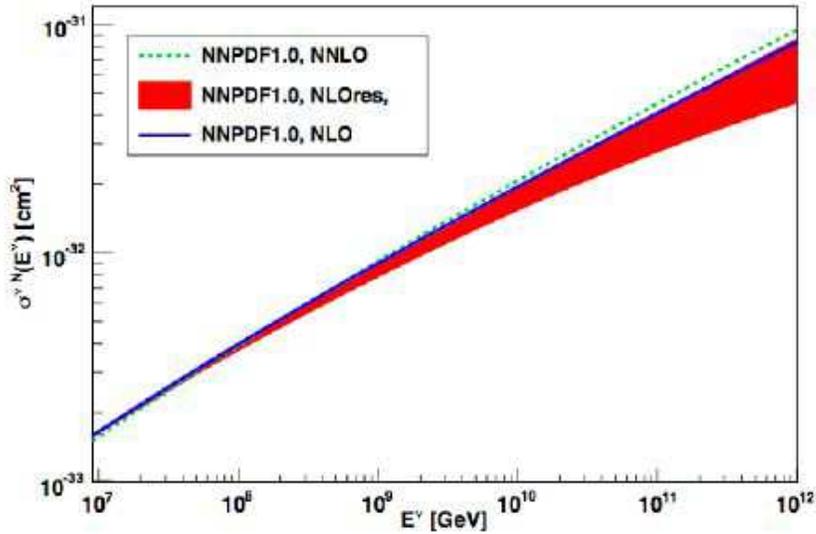} 
\caption[]{The impact of small $x$ resummation on the cross section for very high energy neutrino cosmic rays.}
\label{fignu}
\end{figure}

Remarkable progress in theory has been realised in the resummation of logarithms to all orders. At small or at large values of $x$ (with $Q^2$ large) those terms of higher order in $\alpha_s$ in either the coefficients or the splitting functions which are multiplied by powers of $\log{1/x}$ or $\log{(1-x)}$ eventually become important and should be taken into account. Fortunately the sequences of leading and subleading logs can be evaluated at all orders by special techniques and resummed to all orders. 
The small-x case is particularly important for the interpretation of HERA where the singlet structure function is the dominant channel that generates the sharp rise of the gluon and sea parton densities at small x.
The small $x$ data collected
by HERA can be fitted reasonably well even at the smallest measured values of $x$ by the NLO QCD evolution equations, so that
there is no dramatic evidence in the data for departures (see the talk by S. Reisert). This is surprising also in view of the fact that the NNLO terms in fixed order perturbation theory for the splitting function from the Moch et al calculation  \cite{ver} are quite large. Resummation effects have been shown to resolve this apparent paradox (see the recent review in \cite{heralhc} and refs. therein). For the singlet splitting function the coefficients of all corrections of order $[\alpha_s(Q^2)\log{1/x}]^n$ and $\alpha_s(Q^2)[\alpha_s(Q^2)\log{1/x}]^n$ are explicitly known from the BFKL analysis of virtual gluon-virtual gluon scattering. But the simple addition of these higher order terms to the fixed order perturbative result (with subtraction of all double counting) does not lead to a converging expansion (the NLO logs completely overrule the LO logs in the relevant domain of $x$ and $Q^2$). A sensible expansion is only obtained by a proper treatment of momentum conservation constraints, also using the underlying symmetry of the BFKL kernel under exchange of the two external gluons, and especially, of the running coupling effects . In Fig.~\ref{figx} we present the results for the dominant singlet splitting function $xP(x,\alpha_s(Q^2))$ for $\alpha_s(Q^2) \sim 0.2$. We see that while the NNLO perturbative splitting function sharply deviates from the NLO approximation at small x, the resummed result only shows a moderate dip with respect to the NLO perturbative splitting function in the region of HERA data, and the full effect of the true small x asymptotics is only felt at much smaller values of x. The related effects are moderately small for most processes at the LHC (except for special cases like forward b production) but certainly will become more relevant for next generation hadron colliders. Due to the dip there are less scaling violations at HERA than expected from NLO evolution. At this Workshop J. Rojo has presented K-factors for structure functions at small x for different values of $Q^2$. He has also discussed the results of a resummed calculation of cross sections for ultra high energy neutrinos in cosmic rays (see Fig. (\ref{fignu})). The resummation of small x logs in Drell Yan processes has been accomplished by R. Ball and S. Marzani (presented by the latter). The inclusion of small x effects for the longitudinal structure function has been discussed by A. Stasto. Theoretical aspects of small x resummation have been presented by I. Balitsky, J. Bartels, D. Colferai, V. Fadin, H.P Kowalski and summarised by A. Sabio Vera.   

\section{Unfinished work and open problems}

In spite of the large effort in theory and experiment over about forty years still our knowledge on DIS is in many respects surprisingly not satisfactory. A partial list of examples includes the ambiguities in the extraction of parton density functions from the data, the determination of $\alpha_s$ from DIS, the still poor determination of neutrino structure functions, the fact that only now some reasonable data on the longitudinal structure functions have been presented, obtained by H1 and ZEUS at HERA, the open problems in polarized DIS and so on. We now consider some of these items in more detail.

\subsection{The value of $\alpha_s$ from DIS}

As well known, very precise and reliable measurements of $\alpha_s(m_Z^2)$ have been obtained at $e^+e^-$
colliders, in particular at LEP. The main methods at $e^+e^-$ colliders are: a) The inclusive hadronic Z decay width $\Gamma_h$ that enters in the quantities $R_l$, $\sigma_h$, $\sigma_l$, $\Gamma_Z$ which are independently measured with different systematics. b) Inclusive
hadronic $\tau$ decay. c) Event shapes and jet rates.
The combined value from the measurements at the Z (assuming the validity of the SM and a light Higgs mass) is \cite{klu}:
\beq
\alpha_s(m_Z)=0.1191\pm0.0027\label{alZ}\\
\eeq
The measurement of $\alpha_s(m_Z)$ from $\tau$ decay leads to \cite{dav}:
\beq
\alpha_s(m_Z)=0.1212\pm0.0011\label{altau}\\
\eeq
I am sceptical about the small error quoted in eq. \ref{altau} but this is not an issue here. Important determinations of $\alpha_s(m_Z)$ in $e^+e^-$ annihilation are obtained from different infrared safe observables related to event rates and jet shapes. The main problem of these measurements is the large impact of non perturbative hadronization effects on the result and therefore on the theoretical error. The perturbative part is now known at NNLO. One advantage is that the same measurements can be repeated at different $\sqrt{s}$ values (e.g. with the same detectors at LEP1 or LEP2) allowing for a direct observation of the energy dependence. A state-of-the-art result, from jets and event shapes at LEP is given by \cite{disse}:
\beq
\alpha_s(m_Z)=0.1224\pm 0.0039.\label{aljet}\\
\eeq
Beyond $e^+e^-$ colliders the next golden channel for the determination of $\alpha_s$ should be DIS.
For the determination of $\alpha_s$ the scaling violations of non-singlet structure functions would be ideal,
because of the minimal impact of the choice of input parton densities. 
We can write the non-singlet evolution equations in
the form:
\beq
\frac{d}{dt}logF(x,t)~=~\frac{\alpha_s(t)}{2\pi}\int_x^1\frac{dy}{y}\frac{F(y,t)}{F(x,t)}P_{qq}(\frac{x}{y},\alpha_s(t))
\label{NSEE}\\
\eeq
where $P_{qq}$ is the splitting function  known up to 3-loop accuracy. It is clear from this form that, for example, the normalisation error drops away so that the dependence on the input is reduced to a minimum (indeed, only a single density appears here, while in
general there are quark and gluon densities). Unfortunately the data on non-singlet structure functions are not very
accurate. If we take the difference of data on protons and neutrons, $F_p-F_n$, experimental errors add up in the difference
and finally are large. The $F_{3\nu N}$ data are directly non-singlet but are not very precise. 

A determination of
$\alpha_s$ from the CCFR data on $F_{3\nu N}$ using Mellin moments has led to \cite{kat}:
\beq
\alpha_s(m_Z)=0.119\pm0.006\label{alCCFR}\\
\eeq
Two  independent analyses of the same data using Bernstein moments (a combination of Mellin moments that emphasizes a value of $x$ and a given spread around it in order to be sensitive to the interval where the measured points are) lead to \cite{max} 
\beq
\alpha_s(m_Z)=0.1174\pm0.0043\label{almax}\\
\eeq
but here the theoretical error associated with the method and with the choice adopted for the scale ambiguities, is not considered, and \cite{ynd}:
\beq
\alpha_s(m_Z)=0.1153\pm0.0063\label{alynd}\\
\eeq
Good overall agreement is obtained but the results are not very precise (the realistic error is around $\pm 0.006$) as expected from the limited accuracy of the available neutrino data.

A fit to non singlet structure functions in electro- or muon-production extracted from proton and deuterium data, neglecting sea and gluons at $x>0.3$, was performed in ref.\cite{blum} with the results, at NLO:
\beq
\alpha_s(m_Z)=0.1148\pm0.0019(exp) \pm ?\label{alblu1}\\
\eeq
and at NNLO:
\beq
\alpha_s(m_Z)=0.1134\pm0.0020(exp) \pm ?\label{alblu2}\\
\eeq
Note the rather small central value and that there is not much difference between NLO and NNLO.

When one measures $\alpha_s$ from scaling violations on $F_2$ from e or $\mu$ beams, the data are abundant, the errors
small but there is an increased dependence on input parton densities and especially a strong correlation between the result
on $\alpha_s$ and the input on the gluon density. There are complete and accurate derivations of $\alpha_s$ from
scaling violations in $F_2$. In a well known analysis by Santiago and Yndurain \cite{ynd}, the data on protons from SLAC,
BCDMS, E665 and HERA are used with NLO kernels plus the NNLO first few moments. The analysis uses the Bernstein moments. The quoted result is given by:
\beq
\alpha_s(m_Z)=0.1166\pm0.0013\label{firstalSY}\\
\eeq
A different analysis by Alekhin \cite{ale} of existing data off proton and deuterium targets with NNLO kernels and a more conventional method leads to
\beq
\alpha_s(m_Z)=0.1143\pm 0.0014(exp)\pm 0.0013(th)\label{alSY}\\
\eeq
In both analyses the dominant error is theoretical and, in my opinion, is probably somewhat larger than quoted. The difference in central values between these nominally most precise determinations of $\alpha_s$ suggests a total error of about $\pm 0.003$.

\begin{figure}[!h]
\includegraphics[width=0.8\columnwidth]{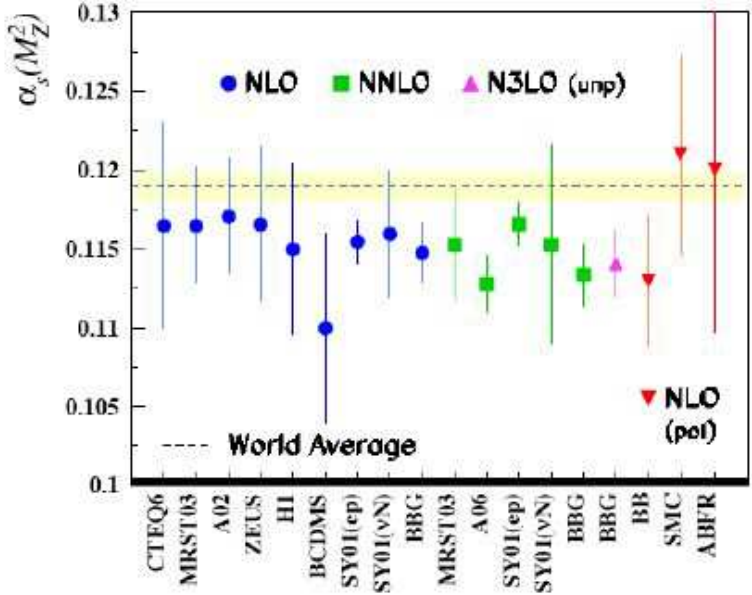} 
\caption[]{The results from the determinations of $\alpha_s$ from DIS compared with the world average (from ref.\cite{blum}).}
\label{alfa}
\end{figure}

An interesting perspective on theoretical errors can be obtained by comparing analyses with different methods. We add the following examples. From truncated moments (but with a limited set of proton data and NLO kernels) \cite{forte}: $\alpha_s(m_Z)=0.122\pm0.006$; from proton data with Nachtmann moments (which take into account some higher twist terms) including effects from soft gluon resummation  \cite{simula}: $\alpha_s(m_Z)=0.1188\pm0.0017 (exp)$. A combination of measurements at HERA by H1 and ZEUS, also including final state jet observables, leads to  $\alpha_s(m_Z)=0.1186\pm0.0051$ \cite{HERA}, most of the error being theoretical.
Finally, to quote a number that appears to me as a good summary of the situation of  $\alpha_s(m_Z)$ from DIS 
one can take the result from a NNLO analysis of available data by the  MRST group \cite{MRST'04} as quoted by \cite{pdg}:
\beq
\alpha_s(m_Z)=0.1167\pm0.004\label{al1SY}\\
\eeq
If we compare all these results on
$\alpha_s$ from DIS with the findings at the Z, given by eq.~(\ref{alZ}-\ref{aljet}), we see that the agreement is good, but the value of $\alpha_s$ from the most precise DIS measurements is somewhat on the low side with respect to $e^+e^-$, as is summarised in Fig.(\ref{alfa}). Finally, we mention the determination of $\alpha_s$ obtained by H1 at HERA from jets in the final state of DIS, presented by A. Specka (related work on jets by ZEUS was reported in the talks by C. Gwenlan and E. Ron):
\beq
\alpha_s(m_Z)=0.1168+0.0050-0.0032\label{al2jet}\\
\eeq
where the error is dominated by the theoretical uncertainties estimated from the scale ambiguities.

\subsection{Ambiguities on Parton Densities}
Much attention has been devoted at this Workshop to the measurements of parton densities, the associated uncertainties and their implications on the forthcoming experiments, notably at the LHC. New analyses and updatings of existing ones have been discussed in the parallel sessions (talks by S. Alekhin,  A. Cooper-Sarkar, H.-L. Lai,  C. Keppel, J. Kretzschmar, P. Nadolsky,  H. Paukkunen,  A. Petrukhin, V. Radescu, J. Rojo, R. Thorne,  M. Ubiali,). Are the parton densities known well enough for our practical purposes? Different fits to the same DIS data lead to parton densities which in general are comparable within errors. But there are important differences from those obtained from all the data \cite{heralhc}. This shows that extrapolation from one data set to another
is still a dangerous operation. An important issue that has been discussed concerns the biases introduced by the assumed parametrizations for parton densities. In this respect the comparison of the results from the fits based on standard parametrizations with those obtained with the qualitatively different procedure based on neural network (NN) methods (discussed by M. Ubiali) is particularly significant. In this approach, an ensemble of parton densities is generated, which pass through randomly generated points within each measurement error bar, and the calculation of measurable quantities is performed by averaging over the ensemble.  The conclusion from this comparison is that the uncertainties found in the unmeasured regions in the NN approach are larger than for the CTEQ, MRST and  Alekhin fits. This is particularly true for the parton densities which are less directly accessible from the data, like the gluon density or, even more, the strange density \cite{NN}.

\subsection{Polarized Deep Inelastic Scattering}

Polarized DIS  is a subject where our knowledge is still far from satisfactory. One main question is how the proton helicity is distributed among quarks, gluons and orbital angular momentum: $1/2\Delta \Sigma + \Delta g + L_z= 1/2$ (talks by E.-C. Aschenauer, S. Bernd, C. Gagliardi, H. Santos, R. Sassot, S. Taneja, F. Taghavi-Shahri; for a recent review, see, for example, \cite{bass}). Experiments have shown that the quark moment $\Delta \Sigma$ is small (the "spin crisis"): values from recent fits give $\Delta \Sigma \sim 0.24-0.34$ (see the talk by M. Stratmann); in any case, a less pronounced crisis than it used to be in the past . From the spin sum rule one obtains that either $\Delta g + L_z$ is relatively large or there are contributions to $\Delta \Sigma$ at very small $x$ outside of the measured region. Denoting, for short hand,  by $\Delta q$ the first moment of the net helicity carried by the sum  $q+\bar q$ we have the relations \cite{strat}:
\beq
a_3= \Delta u -  \Delta d = (F+D)(1+\epsilon _2)=1.269\pm0.003\\
\label{a3}
\eeq
\beq
a_8= \Delta u +  \Delta d -2\Delta s= (3F-D)(1+\epsilon _3)=0.586\pm0.031\\
\label{a8}
\eeq
where the $F$ and $D$ couplings are defined in the SU(3) flavour symmetry limit and $\epsilon_2$ and $\epsilon_3$ describe the SU(2) and SU(3) breakings, respectively. From the measured first moment of the structure function $g_1$ one obtains the value of $a_0=\Delta \Sigma$:
\beq
\Gamma_1= \int dx g_1(x)= \frac{1}{12}\left[a_3+\frac{1}{3}(a_8+4a_0)\right]\\
\label{g1}
\eeq
with the result, at $Q^2\sim 1 \rm{GeV}^2$:
\beq
a_0= \Delta \Sigma=\Delta u +  \Delta d+\Delta s= a_8+3\Delta s \sim 0.24\\
\label{a0}
\eeq
In turn, in the SU(3) limit $\epsilon_2=\epsilon_3=0$, one then obtains:
\beq
\Delta u=0.81,~~~~\Delta d=-0.46,~~~~\Delta s=-0.12\\
\label{uds}
\eeq
This is a strong result! Given $F$, $D$ and $\Gamma_1$ we know $\Delta u$, $\Delta d$, $\Delta s$ and $\Delta \Sigma$ in the SU(3) limit which should be reasonably accurate.
The $x$ distribution of $g_1$ is known down to $x\sim 10^{-4}$ on proton and deuterium and the 1st moment of $g_1$ does not seem to get much at small $x$ (also theoretically $g_1$ should be smooth at small $x$ \cite{gr}). From $\Gamma_1$ on deuterium COMPASS finds $\Delta s = -0.09 \pm 0.01\pm 0.02$, which is consistent with the SU(3) result within errors and given the expected size of possible SU(3) breaking. This is at variance with the value extracted from single particle inclusive DIS (SIDIS) where one obtains
$\Delta s = -0.02 \pm 0.02\pm 0.02$ (presented by H. Santos).

$\Delta \Sigma$ is conserved in perturbation theory at LO (i.e. it does not evolve in $Q^2$). For conserved quantities we would expect that they are the same for constituent and for parton quarks. But actually the conservation of $\Delta \Sigma$ is broken by the axial anomaly and, in fact, in perturbation theory beyond LO the conserved density is actually $\Delta \Sigma'=\Delta \Sigma+n_f/2\pi \alpha_s~\Delta g$  \cite{bass}. Note that also $\alpha_s \Delta g$ is conserved in LO, that is $\Delta g \sim \log{Q^2}$. This behaviour is not controversial but it will take long before the log growth of $\Delta g$ will be confirmed by experiment! But by establishing this behaviour  one would show that the extraction of $\Delta g$ from the data is correct and that the QCD evolution works as expected.  If $\Delta g$ was large enough it could account for the difference between partons ($\Delta \Sigma$) and constituents ( $\Delta \Sigma'$). From the spin sum rule it is clear that the log increase should cancel between $\Delta g$ and  $L_z$. This cancelation is automatic as a consequence of helicity conservation in the basic QCD vertices.  $\Delta g$ can be measured indirectly by scaling violations and directly from asymmetries, e.g. in SIDIS (a beautiful set of data, discussed in the talks by E. Boglione, U. D'Alesio and R. Sassot). Existing measurements by HERMES, COMPASS, and at RHIC are still crude but show no hint of a large $\Delta g$ at accessible values of $x$ and $Q^2$.  Present data are consistent with $\Delta g$ large enough to sizeably contribute to the spin sum rule but there is no indication that $\alpha_s \Delta g$ can explain the difference between constituents and parton quarks. 

The fit to all data presented by M. Stratmann \cite{strat} leads to puzzling results. There is a tension between the 1st moments as determined from the approximate SU(3) symmetry and from fitting the actual data ($x\geq 0.001$) (in particular for the strange density). One could question, on the one hand, the adequacy of the SIDIS data (in particular of the kaon data which fix $\Delta s$) and of their theoretical treatment (for example, the application of parton results at too low an energy) and, on the other hand, the effect of the possibly too rigid parametrization that is assumed in order to determine the polarized densities from the data (remember the above discussion on the ambiguities on parton densities and the comparison with the neural network approach). One finally must keep in mind that the present uncertainties on the gluon and strange densities at small x are still very large. In fact, it is possible that the sum rules are reconciled with the data by terms at very small x (even delta function terms at $x=0$) \cite{forteold}, \cite {bass}.

\section{Is there a Future for Deep Inelastic Scattering Experiments?}

As we have seen many important issues remain open in the physics of DIS which are essential for the understanding of QCD, like the longitudinal structure function, heavy quark structure functions, the small-x behaviour of structure functions, polarized parton densities, generalized structure functions (still in their infancy, see the talks by G. Moreno, S. Liuti, A. Movsisyan, A. Mukherjee,  M. Polyakov) diffraction phenomena and so on. Also the measurements of the strong coupling and of parton densities will remain important as a prerequisite for precision hadron collider physics. Thus there are strong physics motivations for continuing experiments in the area of DIS in the future. The last generation of experiments is near to the end. HERA is now closed although important new results are still being produced from the analysis of the collected data. The CERN experiment COMPASS (presented by A. Magnon) is also approaching its conclusion.  At JLAB ( see the talks by S. Kuhn and R. Ent, M. Guidal, C. Keppel, S. Liuti) one is preparing for the 12 GeV upgraded phase. The main programmes for a future of DIS experiments are, on one side of the Atlantic, EIC, the Electron Ion Collider, (see the talk by  A. Deshpande)
which can take the form of ELIC at JLAB or of eRHIC at Brookhaven, and, on the other side, the long term project LHeC at CERN (A. De Roeck, A. Caldwell, M. Klein). These programmes where also discussed and put in a general perspective at the  LHeC/EIC Panel Discussion. The EIC is a high luminosity ($L \geq 10^{33} \rm{cm}^{-2}s^{-1}$), intermediate energy (c.o.m energy $E=10 - 100 ~\rm{GeV}$) e-A collider  with possibility of polarizing protons and of circulating a heavy ion beam for QCD studies, nuclear phenomena and proton spin physics. The LHeC is an e-p or e-A collider obtained by colliding a $e^{\pm}$ beam from a new accelerator with a LHC beam. The energy parameters are $70 ~\rm{GeV}~ e^{\pm}$ against  $7 ~\rm{TeV} $ protons to give  a c.o.m. energy of $E_{CM}\sim 1.4~ \rm{TeV}$ (compare with HERA $E_{CM}\sim 0.3 \rm{TeV}$). The luminosity goal is $L \sim 10^{33} \rm{cm}^{-2}s^{-1}$, or 3-30$ ~\rm{fb}^{-1}$ per year, (HERA ~ 0.12-0.3 $\rm{fb}^{-1}$ per year). The boost factor $\gamma$ of the ep system is given by $\gamma=E_p/m_{ep}\sim 7/1.4=5$ , compared to $\gamma=2.7$ at HERA. The realization of $e^{\pm}$ polarization is possible as well as simultaneous running of ep with pp or eA with AA. The broad physics goals are the study of proton structure and precision QCD physics in the domain of x and $Q^2$ of LHC experiments, the experimental investigation of small-x physics in ep and eA collisions, probing the $e^{\pm}$-quark system at ~TeV energy in search of leptoquarks, excited electrons, mirror electrons, SUSY with no R-parity and so on, the search for new EW currents (RH weak currents, effective eeqq contact interactions...).
A dedicated Workshop on the LHeC is planned at Divonne next September (it will be its second edition) and the preparation of the Conceptual Design Report is under way. I think that in view of the investment done for the LHC and given that in perspective DIS experiments will remain crucial for the continuation of particle physics, the LHeC is a well motivated option, more so if the next generation $e^+e^-$ machine is built elsewhere than at CERN. Of course, all that becomes realistic if the outcome of the LHC will be rich in discoveries and particle physics will be boosted again.

In conclusion, very interesting data are still coming out from DIS experiments. Many problems and challenging goals remain. The continuation of experiment and theory in this domain is well motivated also in view of the LHC and beyond.

\newpage

\section{Acknowledgments}

It is with great pleasure that, on behalf of all participants, I thank the local Organizers, in particular Juan Terron, who have done a really great job.  The very efficient Palacio de Congresos has been an ideal venue for the Conference and the brilliant town of Madrid offered a very exciting background.

\end{document}